\documentclass[a4paper,english]{aa}
\usepackage{times}
\usepackage[T1]{fontenc}
\usepackage{amsmath}
\usepackage{graphicx}
\usepackage{amssymb}
\usepackage{natbib}
\usepackage{babel}
\usepackage{xspace}
\bibpunct{(}{)}{;}{a}{}{,}
\newcommand{\ascc}   {\mbox{ASCC-2.5}\xspace}

\defcitealias{intpar}{Paper~I}
\defcitealias{clussp}{Paper~II}
\begin{document}

\title{The evolution of luminosity, colour and the mass-to-luminosity ratio of Galactic open clusters:}

\subtitle{comparison of discrete vs. continuous IMF models}

\author{A.E.~Piskunov \inst{1,2,3} \and
        N.V.~Kharchenko \inst{1,3,4} \and
        E.~Schilbach \inst{3} \and
        S.~R\"{o}ser \inst{3} \and
        R.-D.~Scholz \inst{1} \and
        H.~Zinnecker \inst{1} }

\offprints{R.-D.~Scholz}

\institute{Astrophysikalisches Institut Potsdam, An der Sternwarte 16,
D--14482 Potsdam, Germany\\
email: rdscholz@aip.de
\and
Institute of Astronomy of the Russian Acad. Sci., 48 Pyatnitskaya
Str., 109017 Moscow, Russia
\and
Astronomisches Rechen-Institut, M\"{o}nchhofstra\ss{}e 12-14, Zentrum f\"ur Astronomie der
    Universit\"at Heidelberg,
D--69120 Heidelberg, Germany
\and
Main Astronomical Observatory, 27 Academica Zabolotnogo Str., 03680
Kiev, Ukraine
}

\date{Received 12 July 2010 / Accepted 12 October 2010}

\abstract{We found in previous studies that standard Simple Stellar Population (SSP) models are unable to describe or explain the colours of Galactic open clusters both in the visible and in the NIR spectral range. The reason for this disagreement is the continuous nature of the stellar IMF in clusters which is the underlying assumption in the SSP models. In reality, the Galactic open clusters are scarcely populated with the brightest stars responsible for integrated fluxes.}
%aims
{In this study, we aim at constructing discrete SSP-models which are able to adequately describe the observed magnitude-, colour-, and mass-to-luminosity-ratio-age relations of open clusters by including a number of rarely considered effects. }
%methods
{We construct a numerical SSP-model, with an underlying Salpeter IMF, valid within an upper $m_u$ and lower $m_l$ stellar mass range, and with total masses $M_c=10^2\dots10^4\,m_\odot$ typical of open clusters. We assume that the mass loss from a cluster is provided by mass loss from evolved stars and by the dynamical evaporation of low-mass members due to two-body relaxation. The data for the latter process were scaled to the models from high-resolution N-body calculations. We also investigate how a change of the $m_l$-limit influences magnitudes and colours of clusters of a given mass and derive a necessary condition for a luminosity and colour flash. }
%results
{The discreteness of the IMF leads to bursts in magnitude and colour of model clusters at moments when red supergiants or giants appear and then die. The amplitude of the burst depends on the cluster mass and on the spectral range; it is strongly increased in the NIR compared to optical passbands. In the discrete case variations of the parameter $m_l$ are able to substantially change the magnitude-age and $M/L$-age relations. For the colours, the lowering of $m_l$ considerably amplifies the discreteness effect. The influence of dynamical mass loss on colour and magnitude is weak, although it provides a change of the slopes of the considered relations, improving their agreement with observations. For the Galactic open clusters we determined luminosity and tidal mass independent of each other. The derived mass-to-luminosity ratio shows, on average, an increase with cluster age in the optical, but gradually declines with age in the NIR. The observed flash statistics can be used to constrain $m_l$ in open clusters.}
%conclusions
{}

\keywords{
Galaxy: open clusters and associations: general --
Galaxy: stellar content --
Galaxies: fundamental parameters --
Galaxies: photometry --
Galaxies: starburst --
Galaxies: star clusters}

\maketitle

\section{Introduction}
\label{sec:intro}

Star clusters are the best tracers of star formation and evolution of Galactic populations, both in the Milky Way and in external galaxies. In their majority, extragalactic clusters are observed as unresolved objects. To determine their parameters one frequently uses Simple Stellar Population (SSP) models based on methods of evolutionary synthesis. As a benchmark of such models, it is reasonable to use the data on resolved clusters where the physical parameters can be determined independent of the integrated light. Such clusters should be resolved and observable down to faint magnitudes. Obviously, the best candidates for this kind of investigations are open clusters in nearby galaxies or in the Milky Way itself.

We apply this approach to the  local population of  Galactic open clusters. The basic data on open clusters are taken from our previous work. These are cluster ages \citep[see][]{clucat,newclu}, tidal masses \citep{clumart}, and integrated photometric magnitudes (\citealt{intpar}, referred hereafter as Paper~I).

The comparison of the observed cluster colours with the prediction of the standard SSP-models GALEV \citep{galev03} and Starburst99  \citep{sb99_05} revealed significant discrepancies for clusters younger than 1 Gyr. The observed clusters usually are bluer than the models predict, but from time to time they show a sudden ``red colour flash''. Our analysis (\citealt{clussp}, Paper~II) has shown that these discrepancies occur when the discrete nature of the IMF is neglected in the models.

In this paper we consider the properties of SSP-models and try to achieve compatibility of the models with open cluster data by applying the well known approach of stochastic modelling of star clusters \citep[see e.g. ][]{bruz02} adopted in many cluster studies \citep[see e.g.][]{girarea95,cervino04,raimea05,popha10,foulan10}. In order to make our models as close to reality as possible, we include a number of effects which might be important in the evolution of open clusters of low or moderate masses. The paper analyses the evolution of luminosity, colour, and the mass-to-luminosity ratio with cluster age, comparing SSP models based on a continuous IMF with models based on a discrete IMF.

The paper has the following structure. In Sect.~\ref{sec:data} we briefly describe the data on open clusters and adopted definitions. Sect.~\ref{sec:model} gives the recipes for the construction of realistic SSP-models. In Sect.~\ref{sec:lumev}, \ref{sec:colev}, and \ref{sec:ml} we discuss the evolution of the SSP-models in magnitude-age, colour-age, and $M/L$-age diagrams and compare the model results with observations. In Sect.~\ref{sec:flashpop} we derive the condition for producing a luminosity flash in a star cluster and use this condition to derive additional properties of Galactic open clusters. In Sect.~\ref{sec:conc} we summarise the basic results of this study.

%----------------------------------------------------------------------------%
\begin{figure}[t]
\resizebox{0.7\hsize}{!}{
\includegraphics[origin=c,angle=270,clip]{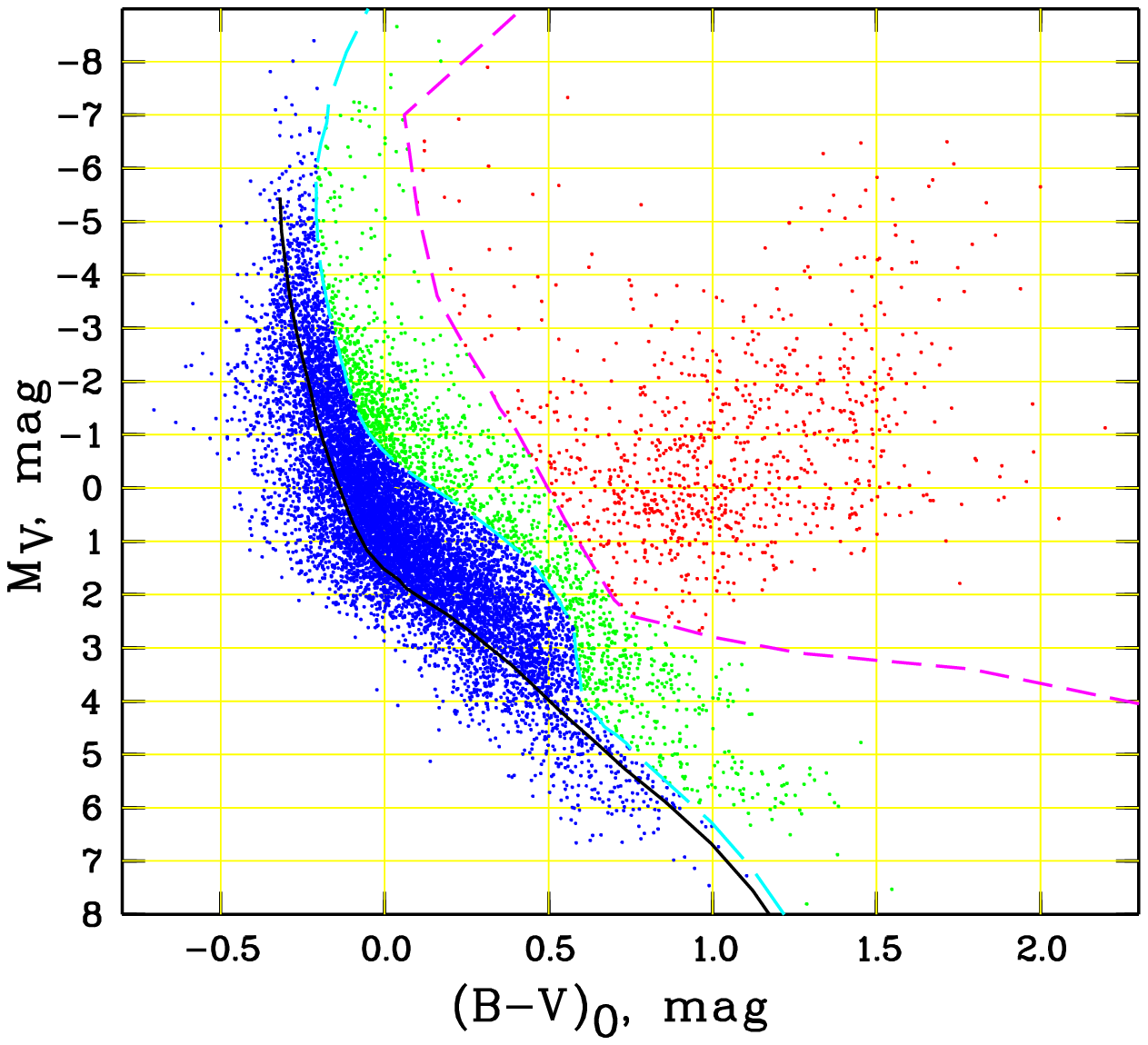}
}\\
\caption{The empirical colour-magnitude diagram based on data from the \ascc of the most probable cluster members used for the classification of 650 open clusters. The solid black curve is the empirical ZAMS of \citet{schkal82}, the solid cyan curve is the TAMS taken from the Padova models, and the broken (magenta) curve separates the Red Giant domain from post-MS stars. The members located to the left of the TAMS are classified as Main Sequence stars (marked in blue colour), those located to the right of the broken line as Red Giants (red), and stars in between (green) are cluster members which just have left the Main Sequence but are not yet red enough (Post-MS stars).
}
\label{fig:cmds}
\end{figure}
%----------------------------------------------------------------------------%

\section{Data and definitions}
\label{sec:data}

%----------------------------------------------------------------------------%
\begin{figure}[t]
\resizebox{\hsize}{!}{
\includegraphics[origin=l,angle=270,clip=]{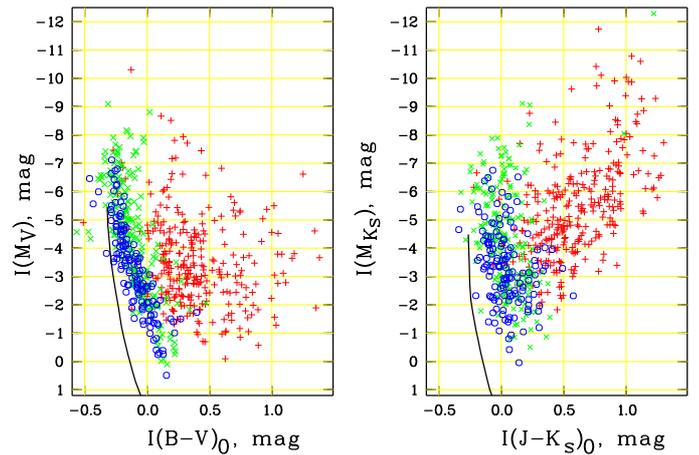}
}
\caption{The empirical colour magnitude diagrams of 650 open clusters identified in the \ascc in the optical (left panel) and the NIR (right panel) photometric passbands. The symbols follow the classification scheme shown in Fig.\ref{fig:cmds}. The blue circles mark the clusters with Main Sequence stars only, the green crosses are those clusters, where Post-MS stars are also found, and the red pluses indicate clusters which contain at least one Red [Super]Giant. For comparison, the solid black curve shows the empirical ZAMS of Schmidt-Kaler for stars.
}
\label{fig:cmdc}
\end{figure}
%----------------------------------------------------------------------------%

In this study, we consider a sample of 650 local open clusters identified in the \ascc\footnote{All-sky Compiled Catalogue of 2.5 million stars, 3rd version, CDS I/280B, \texttt{ftp://cdsarc.u-strasbg.fr/pub/cats/I/280B/}}. In the latest version of the \ascc catalogue, NIR magnitudes have been included which were taken from the 2MASS Point Source Catalogue \citep{cat2MASS}. For each cluster, the basic astrophysical parameters (age, distance etc.) were determined from homogeneous photometric and astrometric data by taking into account the membership probabilities of cluster stars \citep[see][]{clucat,newclu}. The integrated magnitudes were obtained in the optical $BV$, and the NIR $JHK_s$ passbands by summing up the individual luminosities of the most probable cluster members \citepalias[see][ for more details]{intpar}. In order to convert  apparent magnitudes to absolute integrated magnitudes, we used the distances and extinction values given in \citet{clucat,newclu}. The colours were computed as differences of the magnitudes in the corresponding passbands.

From a comparison of our results with independent estimates of integrated magnitudes and colours available in the literature for Galactic open clusters \citep{battin94,lata02}, we found the rms-errors of the differences to be $\pm0.6$ mag and $\pm0.2$ mag for magnitudes and colours, respectively \citepalias{intpar}. These values may be regarded as the upper limits of the uncertainties of our data. The main reason for the differences arises from the different membership criteria applied: in case of solely photometric selection of cluster members around the cluster centre (as usually happens in previous determinations), integrated luminosities can be overestimated due to admixture of bright foreground stars, or underestimated due to missing of relatively bright members outside the central cluster region. However, in the following we assume our data to be accurate within $\pm0.6$ mag for magnitudes and $\pm0.2$ mag for colours though, one can expect that the realistic accuracy is a factor of two better.

In Figs.~\ref{fig:cmds} and \ref{fig:cmdc} we show the combined colour-magnitude diagrams (CMDs) for the most probable members of all our clusters, and the integrated CMDs of the clusters themselves. The figures explain our cluster classification which will be used for convenience throughout the paper. According to the evolutionary status of their brightest stars, we consider three groups of clusters: the Main Sequence (MS) Clusters, the Red Giant (RG) Clusters and Post-MS Clusters. The Main Sequence Clusters contain Main Sequence stars only, the Red Giant group consists of clusters with at least one red giant or supergiant as a member in their upper CMD, and in the Post-MS clusters one observes that the brightest star just left the Main Sequence, but did not yet reach the Red Giant/Supergiant domain. The classification of members into MS, Post-MS or giant stars is based on the data of the Padova isochrones. As separation lines we use the TAMS, and the line connecting the points of maximum luminosity at the shell hydrogen-burning stage. Since stars cross the Hertzsprung gap fast, the exact location of this limit is not so important.

The other data  of relevance for this paper are the cluster masses. They were estimated from the tidal radii of clusters which were drawn either from a direct fitting of King's profiles to the observed density distribution of cluster members \citep{clumart} or from calibrating the apparent semi-major axes of the projected member distribution into the tidal radii system \citep{clumart1}. It must be emphasised that in our approach cluster masses and cluster luminosities are estimated independent of each other. This allows an unbiased construction of the $M/L$ ratio. This circumstance overcomes the general disadvantage of the uncertainty of tidal masses of open clusters and makes their use in this study feasible. 

\section{The construction of the SSP models}
\label{sec:model}

A number of SSP-models has been constructed during the last decade. They are widely used for modelling both star clusters and galactic populations \citep[e.g.][]{mouhcin02,bruz03,maraston05}. In this context we mention a study by \citet{alaba09} who included the effect of mass loss via evaporation of stars from clusters in standard SSP-models, and investigations by \citet{popha10} who take into account the discreteness of the initial mass function (IMF). However, no SSP-models are available incorporating both effects. Since these issues seem to be of great relevance for modelling Galactic open clusters, we refine the model presented in \citetalias{clussp} by considering both effects. In order to make sure that the models we produce are compatible  with present-day SSP models, we construct models both in the ``standard'' and ``extended'' modes. The standard mode reproduces the features of the adopted  SSP models, whereas the extended mode takes into account both the effect of a discrete IMF and the effect of mass loss. In this section we describe the model ingredients and compare the results of our modelling with published standard SSP-models.

\subsection{Star formation}
\label{sec:sf}

The basic model parameter is the cluster mass $M_c$. We assume that at the moment of cluster formation (as a classical open cluster) the placental gas has already been removed and cluster consists of stars only. For simplicity, we also assume a simultaneous formation of stars of different masses. The IMF of the cluster stars formed with mass $m$
\[
f(m)=dN/dm\,
\]
is defined within the lower and upper limits ($m_l$ and $m_u$) of the allowed masses of cluster stars. Here $dN$ is the number of stars having masses between $m,m+d\,m$. We adopt a power-law form for the IMF,
\begin{equation}
f(m)=k\,m^{-\alpha}\, \label{eq:powimf}
\end{equation}
with parameters $\alpha$ (IMF slope), and $k$ (normalisation factor) where $k$ using proper normalisation can be computed via cluster mass $M_c$, slope $\alpha$, and the limits $m_l$ and $m_u$
\[
k = \frac{(2-\alpha)\,M_c}{m_u^{2-\alpha}-m_l^{2-\alpha}}\,.
\]
There are several reasons of choosing a simple power-law IMF valid over the whole range of stellar masses. First of all, the integrated fluxes of open clusters do not depend strongly on the luminosity nor on the number of low-mass stars (i.e. IMF parameters at lower masses). Secondly, the mass spectrum at the low-mass end is changing with time due to dynamical evolution, so the initial distribution of stars is no longer valid for older clusters. Finally, the low-mass IMF is poorly known, and different possibilities (including the  Salpeter IMF) are still under discussion \citep{maps97}. Also, recent observations indicate that the cluster mass spectrum follows a power-law shape with an exponent of about $-1.8$ \citep{bobajea10} down to small masses of $m\sim 0.1\,m_\odot$. Therefore, we decide to restrict the modelling to a power-law IMF confined with some ``effective'' low mass limit $m_l$. Throughout the  paper, the mass spectrum is represented by the Salpeter IMF ($\alpha=2.35$).

The largest possible mass of cluster stars $m_u$ is defined by the adopted grid of evolutionary tracks/isochrones. Only at the very early stages of cluster evolution does $m_u$ have an impact on the integrated colours. In contrast, the lower mass limit $m_l$  governs the population at the massive end of the IMF provided that the IMF slope and the initial cluster mass $M_c$ are fixed.
In the following, we consider $M_c$ and $m_l$ to be free parameters of the adopted model.

\subsection{Stellar evolution}
\label{sec:sev}

The photometric properties of model clusters are defined by the  implemented grid of isochrones. Here we used the grid provided by the Padova group \citep{bertelli94,girarea00,margi07,marigo08} via the online server CMD\footnote{\texttt{http://stev.oapd.inaf.it/cgi-bin/cmd}}. According to our experience, the Padova isochrones fit the brighter parts of the open cluster CMDs reasonably well over a wide range of cluster ages. These parts are responsible for the integrated luminosities and colours. We used a solar metallicity grid ($Z=0.019$), and retrieved the passbands $U,B,V,R,I$ and $J,H,K_s$ with an age range $\log t=6.0\dots10.2$ and a step of 0.02 in $\log t$.

\subsection{Discretization of the model}
\label{sec:rand}

Adopting the continuity of the IMF in standard SSP models, one assumes that every (infinitely) small mass interval of the given isochrone emits light according to its $T_{eff}$ and $\log g$. The total flux that a ``theoretical'' cluster emits is a result of the integration over the isochrone, weighted by the given IMF. It is immediately obvious that in this case luminosity must scale with mass, once the age (or the isochrone) is fixed. In the discrete approach, however, only those loci on the isochrone contribute where stars of a respective mass are actually present.

With the distribution function of stars given by the IMF, we construct an initial sample of cluster members for a given set of model parameters. The total number of members $N_c$ is defined by the integral
\[
N_c = \int\limits_{m_l}^{m_u}f(m)\,dm,
\]
and the mass of the $k$-th member $m_k$ is assigned at random by the distribution given by eq.~(\ref{eq:powimf}). At a given age of a cluster, each cluster member contributes to the cluster luminosity in accordance with its evolutionary state.

For a given $M_c$, open clusters represent random realisations of the IMF which have statistical fluctuations even for a smooth probability density distribution. At the high-mass end of the mass range, these fluctuations are particularly large because of low number statistics, and produce gaps in the mass spectrum. The fluctuations depend not only on the model mass, but also on the lower mass limit which, at a given $M_c$, controls both the upper mass limit and number of massive stars. For example, for the Salpeter IMF, $M_c=1000\,m_\odot$ and $m_l=0.01,\,0.1,\,$ and $1\,m_\odot$, the upper mass limit $m_u$ turns out to be about 60, 90, and $100\,m_\odot$, on average. Note that $m_u$ is confined to $100\,m_\odot$ due to the adopted evolutionary grid which does not include higher mass stars.

%----------------------------------------------------------------------------%
\begin{figure}[t]
%\sidecaption
\resizebox{\hsize}{!}{\includegraphics[clip]{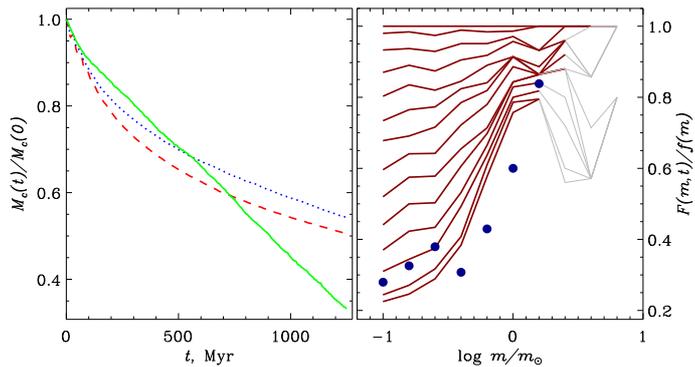}}
\caption{Mass loss from star clusters according to high-resolution N-body models by \citet{bepe08}. The left panel shows the relative loss of the total mass of clusters orbiting at the Solar Galactocentric distance ($R_G=8.5$ kpc). The solid (green), dotted (blue), and dashed (red) curves are for clusters of initial masses $M_c= 10^3,\,5 \cdot 10^3,\,10^4\,m_\odot$, respectively. The right panel demonstrates selective mass loss occurring among stars of different masses $m/m_\odot$ in a cluster with $M_c=10^3\,m_\odot$. The curves show the evolution of the MF-to-IMF ratio $F(m,t)/f(m)$ with time, starting from $t=0$~Myr, with a step of 100~Myr and last point $t=1249$~Myr. The dots show the data of \citet{alaba09} scaled to a cluster with mass $M_c=10^3\,m_\odot$ and the age of 1249 Myr. The red portions of the curves correspond to  stars alive and grey ones to stellar remnants.
}
\label{nbody_fig}
\end{figure}
%----------------------------------------------------------------------------%

\subsection{Mass loss from the cluster}
\label{sec:malo}

Usually the standard SSP-models neglect mass loss from star clusters. Indeed, the bulk of the cluster luminosity comes from the brightest stars. Their mass distribution is relatively weakly affected by dynamical processes if compared to low-mass stars. On the other hand, masses of extragalactic clusters are usually determined from the mass-luminosity relation ($M/L$ ratio) which is also based on the standard SSP models. A recent attempt to incorporate dynamical processes into SSP models was made by \citet{alaba09} who included results of N-body calculation of \citet{bamak03} into a standard GALEV model (i.e., with continuous IMF).

In the present paper, we use the results of N-body simulations by \citet{bepe08} which were carried out to reproduce the local population of the Galactic open clusters\footnote{The full model snapshots are available from ftp://ftp.ari.uni-heidelberg.de/pub/staff/berczik/phi-GRAPE-cluster.}. Below we give a short description of the model parameters. For more details see \citet{elli}. The models represent ensembles of stars distributed at birth along with the Salpeter IMF ($\alpha=2.35,\,m_l=0.08\,M_\odot,\,m_u=8\,M_\odot$) and moving in the external tidal field of the Milky Way at the Solar Galactocentric circular orbit ($R_G=8.5$ kpc).

Such an approach  means that both dynamical- and SSP-models are coordinated with respect to the properties of the stellar population. The adopted set of N-body model parameters is listed in Table~\ref{tbl:nbmod}. In the first three columns of the table we show initial mass, number of stars comprising a model, radius of the model clusters, and the last column indicates their estimated dissolution time. Further, we assume that the phase of the initial mass loss (``infant weight loss'' according to \citealt{weidea07}) is relatively short and is finished with the death of the most massive cluster members ($m>8\,M_\odot$). We trace the evolution of the N-body model within a time interval of 1.2 Gyr. Consequently, in this study we consider the intermediate phases of cluster dynamical evolution occurring between the infant weight loss and the final decay phases. However, this embraces more than 90\% of the observed local cluster population.

In the context of the present study, the most interesting feature of the dynamical models is the steady mass loss appropriate to clusters due to their dynamical evolution and its consequence for the stellar content of clusters. These functions are shown in Fig.~\ref{nbody_fig}. In the left panel we compare the relative change of the total mass of the model during the first 1.2 Gyr of the evolution. Both effects of stellar and dynamical evolution are taken into account. It is assumed that the gas lost by the stars is removed from the system, whereas the stellar remnants remain within the cluster and continue their dynamical evolution as normal stars of the respective mass. The dynamical mass loss has a regular nature and occurs due to two-body interactions within the cluster. No such events as molecular cloud encounters are taken into account, so our results provide lower limits of possible dynamical effects. According to Fig.~\ref{nbody_fig}, the rate of the relative mass loss is higher in massive clusters at the beginning of their evolution, but after about 600 Myr it slows down, whereas the low-mass clusters continue to loose mass with approximately constant rate. At $\log t \approx 9.2$ the low-mass clusters have almost completely decayed. For more massive clusters this process takes considerably longer time (see Table~\ref{tbl:nbmod}).

\begin{table}
\caption{Parameters of adopted N-body models}\label{tbl:nbmod}
\begin{tabular}{cccc}
\hline
$M_c(0)$, $M_\odot$&$N_c(0)$&$R_c(0)$, pc&$t_{dis}$, Myr\\
\hline
 1000 & 4040 &  3 & 1511\\
5000  & 20202&  7 & 4661\\
10000 &40404 & 10 & 7572\\
\hline
\end{tabular} 
\end{table}

The right panel of Fig.~\ref{nbody_fig} shows how effectively stars of different masses evaporate. As a measure of the dissipation process, we use
\begin{equation}
q(m,t) = F(m,t)/f(m)\,,\label{eq:qmt}
\end{equation}
where $f(m)$ is the initial mass function, and $F(m,t)$ is the mass function (MF) of cluster stars at an instant time $t$. The MF is a result of both stellar and dynamical evolution of the cluster population. The former process results in shortening the massive end of $F(m,t)$ and in decreasing $q(m,t)$, whereas the dynamical evolution is more effective in depleting stars at the low-mass end of $F(m,t)$. The input parameters of our SSP models include the possibility of stellar masses outside the mass range of $0.08\,M_\odot$\dots $8\,M_\odot$ adopted for N-body simulations. In these cases, the constructed relations are linearly extrapolated. As the evaporation is more efficient than the expected value from linear extrapolation, our estimates of $q(m,t)$ are too conservative for stars with masses $m<0.08\,m_\odot$. In contrast, this extrapolation is of minor importance for high-mass stars, since the lifetime of stars with $m>8\,m_\odot$ is much shorter than the time of a significant decline of $q(m,t)$ from unity.

For comparison, we plot the results of \citet{alaba09} in Fig.~\ref{nbody_fig} as filled dots. In order to place their data into Fig.~\ref{nbody_fig}, we compute the MF by using formula (6) from \citet{alaba09}.  A normalisation (i.e. the amount of cluster mass loss) is achieved by shifting the resulting function $q$ along the vertical axis to provide the best fit at $t=1.2$ Gyr. In general, the results from both approaches are in  reasonably good agreement, except for a jump in the \citet{alaba09} data which occurs due to a rather rough separation of the data into two mass intervals. Thus, our approach allows a more accurate approximation of the mass loss process along stellar masses.
 
As a ``by-product'' of the N-body simulations we get data allowing to follow the change of the MF slope caused by the dynamical evolution of a cluster alone. Assuming the Salpeter IMF ($\alpha=2.35$), we obtain $\alpha=2.0\pm0.1$ in the mass domain $(0.08,\,1.0)\,m_\odot$ at $\log t=9.5$. This can be compared with the recent finding of a slope of $\alpha=1.8\pm0.1$ derived by \citet{bobajea10} for the Praesepe mass function based on star counts in the same mass domain. The authors explain the flat MF by the possibility that the Praesepe IMF is different from the Salpeter one. The N-body simulations show, however, that  about the same flattening of the Salpeter IMF can be achieved if dynamical evolution of clusters is taken into account.

\subsection{Integrated flux calculation}
\label{sec:ifc}

For the case of a continuous IMF and simultaneous star formation, the integrated luminosity of a cluster in a passband $i$ at a moment $t$ is computed as
\[
L_i(t) = \int\limits_{m_l}^{m(t)} F(m,t)\,l_i(m,t)\,dm\,,
\]
where $F(m,t)$ is the PDMF, $l_i(m,t)$ is the mass-luminosity relation for stars in the $i$th passband valid at the moment $t$, and $m(t)$ is the mass of stars just finishing their life at the moment $t$. Both functions are taken from the theoretical isochrones described above.

In the case of a discrete IMF, we simply reproduce the calculation of integrated fluxes of real clusters by adding the luminosities of non-evaporated individual stars
\[
L_i(t) = \sum\limits_{k=1}^{N_c(t)} l_i(m_k,t)\,,
\]
the stars are labelled with their initial masses $m_k$, and $N_c(t)$ is the number of cluster members with lifetimes exceeding $t$.

The integrated magnitudes $I(M_i)$ are computed as
\[
I(M_i) = -2.5\, \log L_i - M_i^\odot\,,
\]
with $M_i^\odot$ to be the absolute magnitude of the Sun in the passband $i$. In the case of two passbands ($i$ and $j$), the integrated colour $I(i-j)$ is determined as
\[
I(i-j) = -2.5\, \log L_i/L_j\,.
\]

%----------------------------------------------------------------------------%
\begin{figure}[t]
%\sidecaption
\resizebox{\hsize}{!}{\includegraphics[clip]{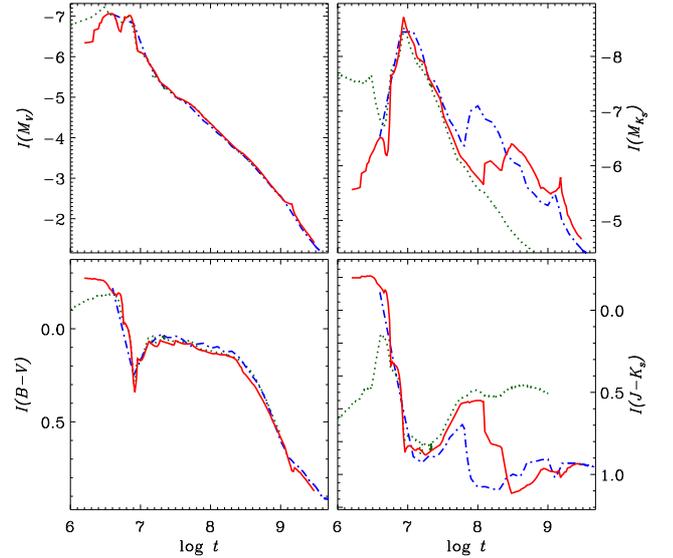}}
\caption{Evolutionary tracks developed in this paper (solid red curves) and models computed with help of the online SSP-servers SB99 (dotted green curve) and GALEV (dash-dotted blue curve). The left panels show optical magnitudes and colours, the right panels are for NIR passbands. The adopted model parameters are given in the text.
}
\label{ssp_fig}
\end{figure}
%----------------------------------------------------------------------------%

\subsection{Comparison with literature models}
\label{sec:compstand}

In Fig.~\ref{ssp_fig} we compare our discrete model with the data computed with the help of online servers provided by the GALEV\footnote{\texttt{http://www.galev.org/}}, and Starburst99\footnote{\texttt{http://www.stsci.edu/science/starburst99/}}(SB99) sites which \textit{de facto} can be regarded as present-day standards. Our model was calculated for a continuous IMF for a cluster with  $M_c=10^3\,m_\odot$, and stellar mass limits $m_l=0.1\,m_\odot$ and $m_u=93\,m_\odot$. The mass spectrum is represented by the Salpeter IMF ($\alpha=2.35$). The on-line continuous models were computed for $M_c=10^6\,m_\odot$, the Salpeter IMF, and stellar mass limits of $m_l=0.1\,m_\odot$ and $m_u=100\,m_\odot$, which  are compatible with our values. We consider the photometric passbands $B,V$ (Johnson) and $JHK_s$ (2MASS). Due to the difference in cluster masses between our and the on-line models, we simply moved the on-line tracks  along the magnitude axis to reach the best agreement between the curves. Since the integrated colours do not depend on cluster masses in the case of continuous IMF, the tracks  in the colour-age diagram are plotted without any shift.

%----------------------------------------------------------------------------%
\begin{figure*}
\resizebox{\hsize}{!}{\includegraphics{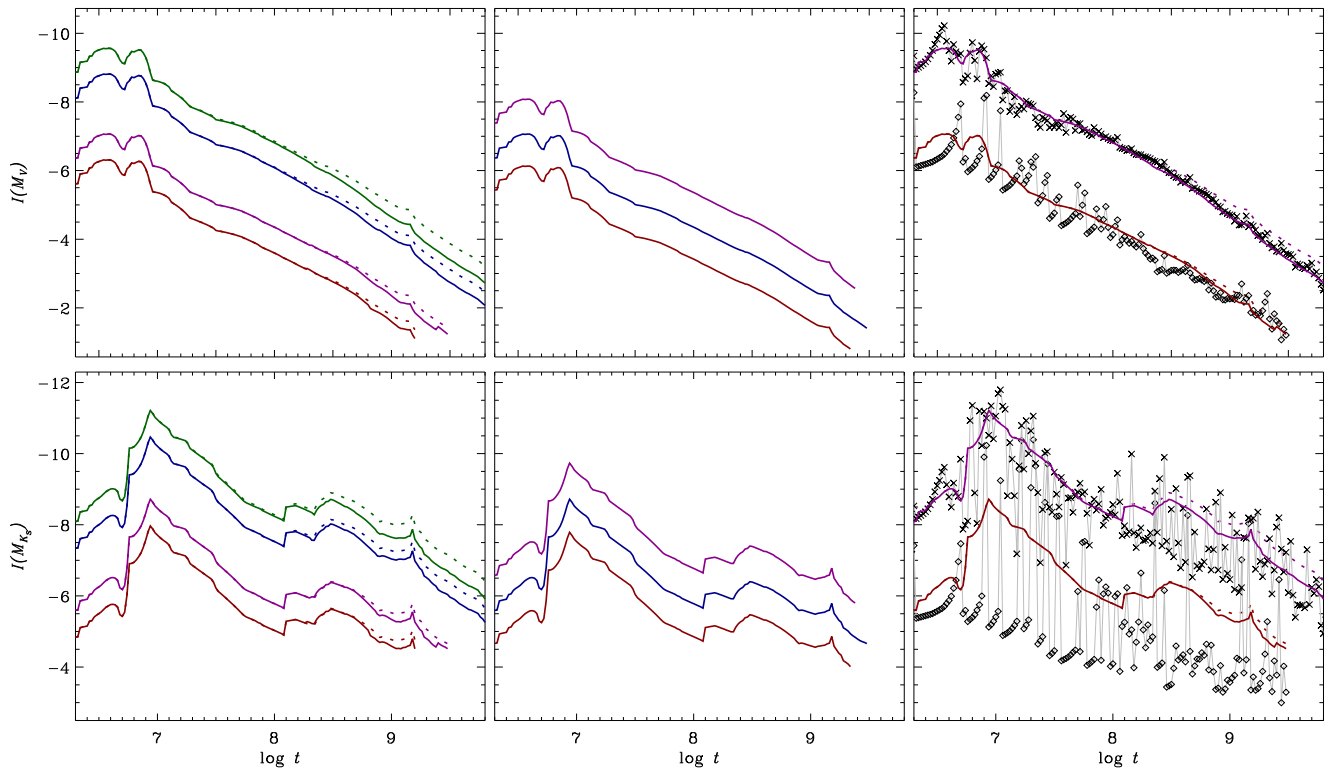}}
\caption{The evolution of cluster luminosity in the optical (upper row) and NIR (lower row). The curves in the left and middle panels present SSP models with continuous IMF, whereas in the right panel we compare continuous and discrete options. \textbf{Left panels:} impact of dynamical  mass loss on integrated luminosity of clusters of different masses and stellar low-mass limit $m_l=0.1\,m_\odot$. The four curves from the top to the bottom are models for clusters with masses $M_c=10^4\,m_\odot$, $M_c=5\cdot 10^3\,m_\odot$, $M_c=10^3\,m_\odot$, and $M_c=5\cdot 10^2\,m_\odot$, respectively. The dotted curves indicate models without mass loss. \textbf{Middle panels:} influence of different lower mass limits $m_l$ of cluster stars on integrated luminosity of clusters with mass $M_c=10^3\,m_\odot$. The upper, middle, and bottom curves correspond to $m_l=1,\,0.1,\,0.01\,m_\odot$, respectively. \textbf{Right panels:} integrated magnitudes of clusters modelled with continuous (curves) and discrete (connected symbols) IMFs for stellar low-mass limit $m_l=0.1\,m_\odot$. The upper curve and crosses are for cluster mass $M_c=10^4\,m_\odot$, and the bottom curve and diamonds are for $M_c=10^3\,m_\odot$. The mass loss due to dynamical evolution is incorporated in both models. For comparison, the continuous model without mass loss is shown by the dotted curves.
}
\label{magef_fig}
\end{figure*}
%----------------------------------------------------------------------------%

According to Fig.~\ref{ssp_fig} (left panels), the models  coincide well in the optical. A small disagreement between our and the SB99 tracks at early ages can be explained by slightly different upper mass limits adopted. In the NIR the agreement between the three models is worse. Especially, the SB99 models differ prominently from both GALEV and our results. This can occur due to a different grid of the underlying isochrones used by SB99. Smaller differences between our models and GALEV probably come from discrepant transformations into the 2MASS system applied in GALEV and our models. A general impression from Fig.~\ref{ssp_fig} is that our models are located in between the data of SB99 and GALEV. Therefore, we conclude that our models describe the same system as the other present-day SSPs.

\section{Luminosity evolution of star clusters}
\label{sec:lumev}

Integrated luminosities (magnitudes) are the basic data providing estimates of masses of extragalactic clusters. They are, also, used to describe the cluster population as a whole, e.g. the history of cluster formation, and its spatial variations (via the luminosity or mass functions). It is generally accepted that the current SSP-models reproduce well the integrated magnitudes of extragalactic clusters. However, there is a number of effects which must be taken into account to avoid serious aberrations by the interpretation of observations.

\subsection{Effects in the magnitude vs. age diagram}
\label{sec:magef}

In Fig.~\ref{magef_fig} (left panels) we check the impact of the mass loss effect on integrated magnitudes of clusters of different masses. We consider models based on the continuous IMF. Within the range of cluster masses from $5 \cdot 10^2\,m_\odot$ to $10^4\,m_\odot$ -- typical for Galactic open clusters -- the spread of cluster luminosities reaches up to three magnitudes both in the optical, and in the NIR. The mass loss due to dynamical evolution leads to some decrease of the integrated luminosity provoked by a relatively minor change of the upper mass spectrum (if compared to the low-mass portion of the IMF). The more massive a cluster is, the earlier the mass loss effect becomes significant. However, the effect is small, from $\Delta M \approx 0.2$ for $M_c < 10^3\,M_\odot$ to $\Delta M \approx 0.5$ for $M_c=10^4\,M_\odot$. This can be compared with the decrease of integrated magnitude due to stellar evolution which reaches about 5 to 6 mag in the optical and about 3 to 5 mag in the NIR.

The impact of the lower mass limit $m_l$ in the mass spectrum of a cluster is shown in the middle panels of Fig.~\ref{magef_fig}. It may seem to be strange that the population of the low-mass end of the mass function can affect the cluster integrated luminosity which is mainly defined by the most massive stars. For the given IMF and cluster mass, however, the lower mass limit governs the position of the ``mass centre'' of the system and hence, the population of the upper part  of the mass spectrum, too. For other IMF shapes  \citep[say][]{misca79}, the low mass limit can be translated into the effective mass limit where the IMF is at its maximum. According to Fig.~\ref{magef_fig}, the effect produces variations of $\Delta M_V\approx 2$ mag when  $m_l$ varies between 0.01 $m_\odot$ and 1 $m_\odot$, and it has the same effect both in the optical and in the NIR.  
 
%----------------------------------------------------------------------------%
\begin{figure*}[t]
\sidecaption
\resizebox{12cm}{!}{\includegraphics[clip]{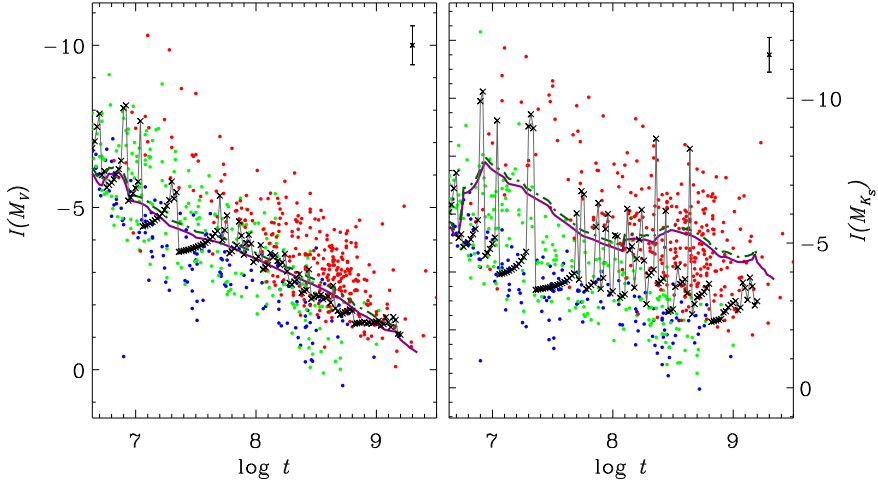}}
\caption{Integrated magnitudes of open clusters in two different passbands vs. cluster age. The observations of the Galactic open clusters are shown by dots, blue for MS-clusters, green for Post-MS clusters, and red for RG-clusters (see Sect.~\ref{sec:data} for explanation). Bars in the upper-right corner of each panel indicate typical errors of the integrated magnitudes. The curves are evolutionary tracks of a continuous model with effects of mass loss incorporated for model parameters $M_c=10^3\,m_\odot$, and $m_l=0.01\,m_\odot$ (solid curve), and $M_c=5\cdot 10^2\,m_\odot$, $m_l=0.1\,m_\odot$ (dash-dotted curve). The latter option ($M_c=5\cdot 10^2\,m_\odot$, $m_l=0.1\,m_\odot$), but with discrete IMF, is shown by the connected crosses.
}
\label{magt_fig}
\end{figure*}
%----------------------------------------------------------------------------%

In the right panels of Fig.~\ref{magef_fig} we compare the influence of the  continuous and discrete IMFs on the integrated magnitudes of open clusters of different masses, and the same $m_l$. In the optical, one can observe a reasonable agreement over the whole age range considered. As expected, the discrete model produces a spread around the continuous track.
At large $M_c$, the discrete option converges to the continuous curve, and at $M_c=10^6$  the discrete and continuous models completely coincide. However, at smaller $M_c$ one observes a systematic difference between integrated magnitudes modelled with the continuous or the discrete IMF.  At short time scales, the track of the discrete model is represented by recurrent events consisting of a slow increase of cluster luminosity due to Main Sequence evolution of the brightest star and of a relatively fast outburst caused by the formation of a bright short-living supergiant.  After the supergiant's decline, the process repeats caused by the follow-up brightest star on the Main Sequence. With increasing cluster age, the outbursts become wider and less intensive due to the lower masses of the stars involved.
     
This scenario is valid for all passbands, but the effects become stronger with increasing wavelengths. The systematics are especially prominent in $K_s$ (see the lower right panel of Fig.~\ref{magef_fig}). Depending on mass and age of clusters, the shift between the models with continuous and discrete IMF reaches from 1 to 3 mag. For younger clusters ($\log t < 7.5$), one is not able to discriminate between the discrete models for $M_c=10^3$ and $10^4\,m_\odot$ in Fig.~\ref{magef_fig}. We attribute this bias to the considerable increase of the luminosity of red supergiants in the NIR with respect to coeval MS-stars. 

\subsection{Comparison with observations}
\label{sec:magefcom}

In Fig.~\ref{magt_fig} we compare the observed magnitude-age relations obtained in different passbands for 650 Galactic open clusters with the corresponding models. The typical accuracy of the derived magnitudes of about 0.6 mag is plotted in the upper right corner of each panel. The observed spread of the data points considerably exceeds this value. The obvious reason is a difference in the initial mass of the observed clusters. However, as we will discuss below, this might not be the sole effect responsible for the scattering. The blue dots indicate the MS-clusters outlining the lower border of the distribution, whereas the red dots (RG-clusters) tend to occupy the upper stripe, and especially the upper-right part of the distribution. This distribution reflects evolutionary changes in the cluster population: the older a cluster is the higher is the probability to observe a red (super)giant among its brightest members. At redder passbands, the distribution becomes more dishevelled though, the segregation between MS- and RG-clusters gets more evident. This points to the relatively higher luminosity of red giants in the NIR.

The observed diagram can be fitted by a set of models with different cluster parameters. In general, the upper bound of the distribution is compatible with cluster masses of the order of $10^4\,m_\odot$, whereas the lower bound can be described by models with a few hundreds of solar masses. In the middle, the cluster models for about $1000\,m_\odot$ can fit the distribution. In Fig.~\ref{magt_fig} we include the models with continuous IMF which fit the middle part of the observed magnitude-age relations. The model parameters chosen are $(M_c,m_l)=(5\cdot 10^2,\,0.1)\,m_\odot$ and $(10^3,\,0.01)\,m_\odot$. An additional scattering of the modelled points can be produced for clusters with $M_c=5\cdot 10^2\,m_\odot$ by varying $m_l$ within the extreme range from 0.01 to $1\,m_\odot$. However, the large spread of  observed integrated magnitudes can be, in a more natural way, described by the models with a discrete IMF.

This can be especially well seen at young ages ($\log t<8$) where open clusters of the same age and the same mass (e.g., $M_c=5\cdot 10^2\,m_\odot$) can have integrated luminosities differing by about 2 mag in the optical and about 5 mag in the NIR. The MS-models with the discrete IMF are systematically located below the corresponding continuous models, but they coincide well with observations of the MS-clusters. During  RG-events, the discrete models  achieve considerably larger luminosities than the values predicted by the standard SSP models but, again, compatible with observed integrated magnitudes of RG-clusters. This has important consequences for the mass determination of open clusters when based on standard SSP models: even perfectly determined integrated magnitudes can provide incorrect cluster masses. Since RG-events are relatively rare and short at earlier stages of a cluster's life (about 20 events during the first 100 Myr for a cluster  with $M_c,m_l = 10^3,\,0.08\,m_\odot$), the mass of this cluster will be, with a higher probability, underestimated by standard SSP models due to the magnitude bias. In the case of RG-clusters, the masses would be considerably overestimated. Again, this conclusion is valid for typical Galactic clusters in the solar neighbourhood. At high cluster masses the discreteness effect gradually vanishes. However, we must keep in mind that the integrated magnitude of a cluster of moderate mass (e.g. $\approx 10^3\,m_\odot$) increases significantly during  RG-events. So, it can be more easily detected in the NIR in other galaxies and assigned an overestimated mass from the standard SSP model.

\section{Colour evolution of star clusters}
\label{sec:colev}

In \citetalias{clussp}, we have already reported on the consequences of using standard SSP models for estimates of cluster ages from observed integrated colours when the effect of a discrete IMF is neglected. In this paper we extend this analysis and consider all the effects studied in the previous sections (mass loss from a cluster, the IMF-discreteness, and the effect of the low-mass limit) and investigate their impact on cluster colours.

\subsection{Effects in the colour-age diagram}
\label{sec:colef}

\begin{figure}
\resizebox{\hsize}{!}{\includegraphics{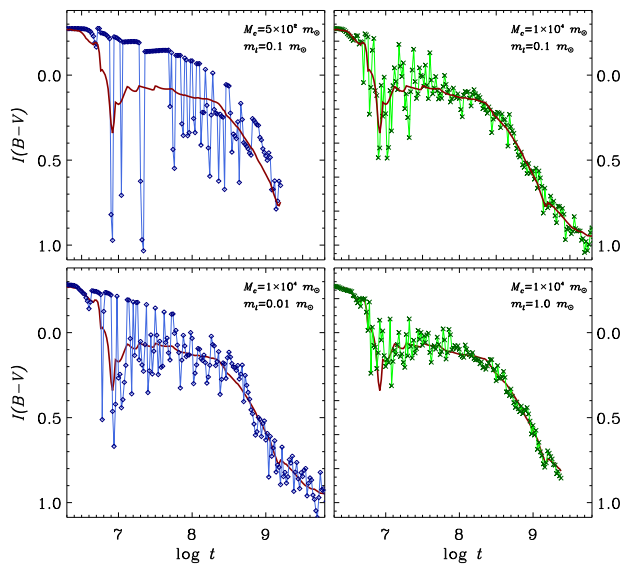}}
\caption{The evolution of cluster colour in the optical regime and the effects of $M_c$ (upper row) and $m_l$ (bottom row). The curve in all panels shows the continuous-IMF model, the connected symbols show the discrete-IMF case. In the upper row we compare $M_c=5\cdot 10^2\,m_\odot$ (left) and $10^4\,m_\odot$ (right) cases for $m_l=0.1\,m_\odot$. The lower row is constructed for $M_c=10^4\,m_\odot$ and $m_l=0.01\,m_\odot$ (left) and $1\,m_\odot$ (right).}
\label{colef1_fig}
\end{figure}

\begin{figure}
\resizebox{\hsize}{!}{\includegraphics{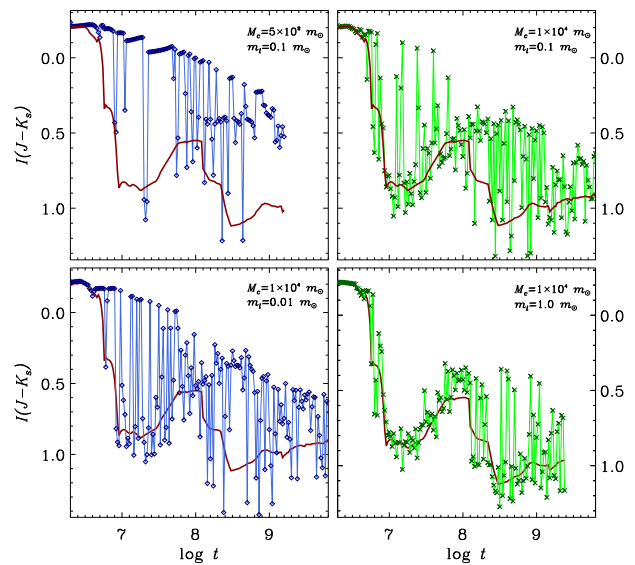}}
\caption{The evolution of cluster colour in the NIR and the effects of $M_c$ (upper row) and $m_l$ (bottom row). The model parameters and designations are the same as in Fig~\ref{colef1_fig}.
}
\label{colef2_fig}
\end{figure}

In Figs.~\ref{colef1_fig} and \ref{colef2_fig} we compare the impact of cluster mass and of the lower limit of stellar masses onto the evolution of integrated colours of a cluster in the continuous and discrete approximations. Again, we consider both optical (Fig.~\ref{colef1_fig}), and NIR (Fig.~\ref{colef2_fig}) colours. As expected, the continuous model does neither depend on $M_c$ nor on $m_l$. The only parameter influencing the standard model is the IMF slope $\alpha$, although this influence is not strong. To substantially change the track, one would need to increase the exponent $\alpha$ to 4.35. Also, a flattening of the IMF with respect to the Salpeter value of the slope does not have a strong effect on the colour-age relation. On the other hand, the discrete model demonstrates a strong dependence on the value of $M_c$ or $m_l$ and shows a considerable divergence from the model with the continuous IMF.

%----------------------------------------------------------------------------%
%----------------------------------------------------------------------------%

First of all, we observe a systematic bias produced due to the discreteness effect. At small masses $M_c$ red giants are rare, and the clusters spend a relatively long time as Main Sequence clusters free of red giants (similar to the present-day Pleiades). In contrast, the continuous models cannot resemble MS clusters (and the Pleiades) since they always contain the RG-portions of the isochrones. This produces a bias which is especially well seen in Fig.~\ref{colef1_fig}, at $M_c=5\cdot 10^2\,m_\odot$ (in the upper-left panel) where we practically observe two different sequences corresponding to the continuous and discrete options. From time-to-time the discrete track is interrupted by a ``red-giant event'', when a cluster becomes extremely red. Simultaneously, as we did see in the previous section, the cluster peaks in its luminosity, so we call such an event a ``red flash''. At $\log t \gtrsim 9$, when the cluster red giants become more long-living and the travelling distance from the Main Sequence to the Rad Giant branch becomes shorter (i.e., the Hertzsprung gap becomes narrower), the sequences converge, and the colour fluctuations diminish. 

With increasing cluster mass the existing holes in the mass spectrum are filled with stars, the IMF approaches a continuous shape, the cluster track approximates the continuous option, and the colour bias is vanishing \citepalias[see][ for a more detailed discussion of this effect]{clussp}. However, the upper-right panel of Fig.~\ref{colef1_fig} shows that at $M_c=10^4\,m_\odot$ -- the upper limit of cluster masses observed in the Solar vicinity -- a sufficient colour bias of the order of $\Delta (B-V)=0.1$ mag is still seen and accompanied by appreciable scatter in colours. Actually, the fluctuations of the colour track reflect the degree of discreteness of the IMF.

A comparison of the upper and lower rows of Fig.~\ref{colef1_fig} also shows that both $M_c$- and $m_l$-effects have approximately the same strengths. The bottom-left plot indicates that for a high-mass cluster the decrease of $m_l$ partly compensates (with respect to the colour bias and fluctuations) the increase of $M_c$. One should note, however, that the colour fluctuations in this case are somewhat smaller, and cease earlier than in the low-mass model with ``normal'' $m_l=0.1\,m_\odot$.

The flashing activity depends on cluster age, but this dependence is different for different cluster masses. Massive flashing clusters are better represented at younger ages ($\log t<7.8$), whereas low-mass clusters show flashes typically later (cf. Fig.~\ref{colef1_fig}). However, the low frequency of ``red-giant events'' in a low-mass cluster is compensated by the large number of low-mass clusters due to the  increase of the cluster initial mass function to lower $M_c$. Therefore, at younger ages one observes both massive and low-mass flashing clusters. At $\log t>7.8$ the massive clusters approach the continuous track, and the flashing population is represented exclusively by low-mass objects.

%----------------------------------------------------------------------------%
%----------------------------------------------------------------------------%

As for integrated magnitudes, the effects for integrated colours become stronger at longer wavelengths. This statement is supported by Fig.~\ref{colef2_fig}, the counterpart of Fig.~\ref{colef1_fig} for $(J-K_s)$ colour. In the NIR we observe strong fluctuations of the order of 1 mag, and a bias of about a few tenths or more even in models of $M_c=10^4\,m_\odot$. The models with discrete IMF always include the effect of mass loss. In the case of a continuous IMF, this effect turns out to be negligible, as the tracks with or without mass loss coincide, in practice.

\subsection{Comparison with observations}
\label{sec:colefcom}

\begin{figure}[t]
%\sidecaption
\resizebox{\hsize}{!}{\includegraphics[clip]{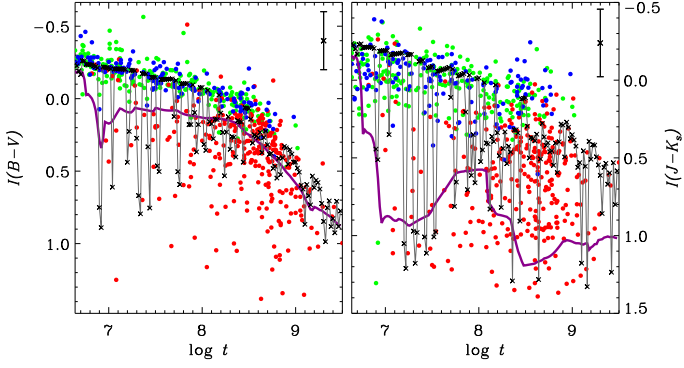}}
\caption{Integrated colours of open clusters vs. cluster age. The left panel shows data for $(B-V)$, right one for $(J-K_s)$. The observations of the Galactic open clusters are indicated by dots, blue for MS-clusters, green for Post-MS clusters, and red for RG-clusters (see Sect.~\ref{sec:data} for explanation). Bars in the upper-right corner of each panel mark typical errors  of the integrated colours. The solid curves are evolutionary tracks of a SSP model with a continuous IMF for a cluster with $M_c=10^3\,m_\odot$ and $m_l=0.1\,m_\odot$. The connected symbols (crosses) show the discrete-IMF option of this model.
}
\label{colobs_fig}
\end{figure}

In Fig.~\ref{colobs_fig} we compare the observed colours of the Galactic clusters  with the adequate models in the colour-age diagram. The observed clusters with ages $\log t \lesssim 8.5$ are predominantly represented by MS-clusters which show a relatively narrow distribution in colour. A redward spread of the data points in the diagram is caused by a few clusters containing yellow or red supergiants, and it is considerably exceeding typical  photometric errors.

At $\log t > 8.5$, one finds, in general, clusters with red giants. The scatter in the observed colours of these clusters is comparable with what is observed for younger ages (opposite to the predictions by the model with $M_c=10^3\,m_\odot$). Part of this scatter can be explained by the different evolutionary stage of red giants along the RG-branch in different low-mass clusters. The other part follows from the lower accuracy of their observations. The older clusters are, on average, absolutely fainter than the younger ones. At distances typical for our sample (<~1kpc), they have fainter apparent magnitudes, sometimes close to the limit of the accuracy of our data. As a consequence, the accuracy of the cluster parameters is typically lower for older clusters than for younger ones. Especially, the lower accuracy of the reddening determination contributes to the colour spread observed for older clusters. The depletion of the cluster sequence after $\log t\approx8.5$ can be explained by either cluster dissolution \citep[the typical lifetime of local clusters is only around $\log t\approx8.5$, according to ][]{clupop} or incompleteness effects, which are more significant for older clusters.

%----------------------------------------------------------------------------%
%----------------------------------------------------------------------------%
 
Again, the model with a continuous IMF is shown in the plots as the solid curve, whereas the discrete option is represented by connected symbols. There is no way for the continuous model to fit the observed colours of either MS-clusters or red ``outliers'', undergoing the red flash. In contrast, the discrete model is able to describe the observed distribution of both groups. The colour scatter of MS-clusters probably reflects the true error of cluster colour determination, as well as the secular colour evolution of the Main-Sequence stars.

In Fig.~\ref{colobs_fig} we show the discrete model with $M_c=10^3\,m_\odot$ and $m_l=0.1\,m_\odot$. It produces a mass spectrum consisting of 2850 stars. In the mass range $5\,m_\odot>m>0.3\,m_\odot$, however, the number of stars is only 625. This is consistent with 780 Pleiades members observed in this mass range by \citet{ple98}. Thus, the model appears to be comparable to the Pleiades. According to the IMF normalisation, the number of model stars with $m>5\,m_\odot$ turns out to be 17. So, before the model cluster reaches the age of the  Pleiades ($\log t\approx 8.1$), it has produced 17 red supergiants. The comparable number of corresponding ``RG-events'' is clearly seen in Fig.~\ref{colobs_fig}. Outside these short events, the cluster is seen as  a main-sequence cluster during the first 125 Myr and resides in the MS-domain of the diagram. If we increase the lower mass limit $m_l$, the number of massive stars increases and the fraction of RG-events slowly increases, especially at older ages. But even at $m_l=1\,m_\odot$, the MS-clusters are present in the track up to $\log t \approx 8.2$. On the other hand, decreasing $m_l$ diminishes the number of RG-events. In principle, from the frequency of red outliers in the ``colour versus $\log t$\,'' diagram, one can estimate indirectly the parameter $m_l$ (see Sect.~\ref{sec:flashpop} more detailed discussion of this issue).

For the other colours the behaviour is qualitatively similar, although the redder the passband, the stronger is the disagreement between continuous and discrete SSPs. As Fig.~\ref{colobs_fig} shows in the $(J-K_s)$ panel, the colour offset between the prediction of the continuous-IMF model and observations reaches  about 1~mag at $\log t<8$. At older ages it tends to be smaller, but still has a significant value of about 0.5 mag. This occurs because the red portions of the isochrone radiate more of their light at longer wavelengths than the blue portions, and the inherent bias increases. Similar to the integrated magnitudes, the higher the cluster mass the smaller is the bias.

%----------------------------------------------------------------------------%
\begin{figure}[t]
\resizebox{\hsize}{!}{\includegraphics{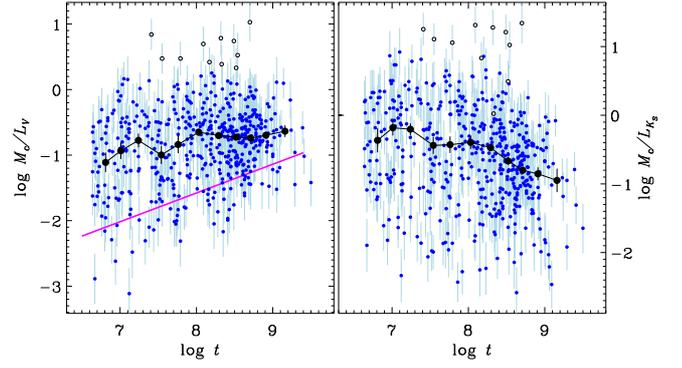}}
\caption{The observed mass-luminosity ratio vs. age of the Galactic open clusters in the optical ($V$) and NIR ($K_s$) photometric passbands. The dots are data for individual clusters, the bars indicate the individual errors in the data. The open dots mark extended clusters excluded from the analysis (see text). The large black dots show the running median of the $M/L$ ratio computed with a ($\log t$)-bin of 0.5 and a step of 0.25. The straight line in the left panel is the empirical relation from \citet{battin94}.
}
\label{mlobs_fig}
\end{figure}
%----------------------------------------------------------------------------%

The observed dispersion in the colour distribution of clusters at $\log t<8.5$ in Fig.~\ref{colobs_fig} can be understood if one assumes the presence of a scatter of initial masses among the observed clusters. The models for low- and medium-mass clusters ($M_c\lesssim10^3\,m_\odot$) provide the MS-clusters and red-outliers. The models for more massive clusters (with $M_c$ about and more than $10^4\,m_\odot$) provide moderately red clusters in the vicinity of the continuous track. At older ages the tracks with continuous and discrete IMF converge. According to Fig.~\ref{colobs_fig}, we conclude that the Galactic open clusters build a sequence of MS-clusters and do not show a concentration towards  the model with a continuous IMF. This agrees qualitatively with a power-law mass distribution of the clusters \citep[see e.g.][]{fuma} which implies the dominance of low-mass clusters in the Solar neighbourhood.

\section{Mass/luminosity ratio -- age relation}
\label{sec:ml}

The mass-luminosity ratio $M/L$ plays an extremely important role in studies of extragalactic clusters. Unfortunately, it is established empirically for massive star clusters like globulars and almost completely unknown for less massive clusters. The major reason has been the low number of reliable masses of open clusters. That is why the mass-luminosity ratio is normally taken either from counted masses or from model predictions. The models themselves need to be constrained by the observations. In this section, we construct the observed $M/L$-age relations based on tidal masses of open clusters and use them for constraining SSP-models, where we take into account several effects usually neglected as model ingredients.

%----------------------------------------------------------------------------%
\begin{figure*}[t]
\resizebox{\hsize}{!}{\includegraphics{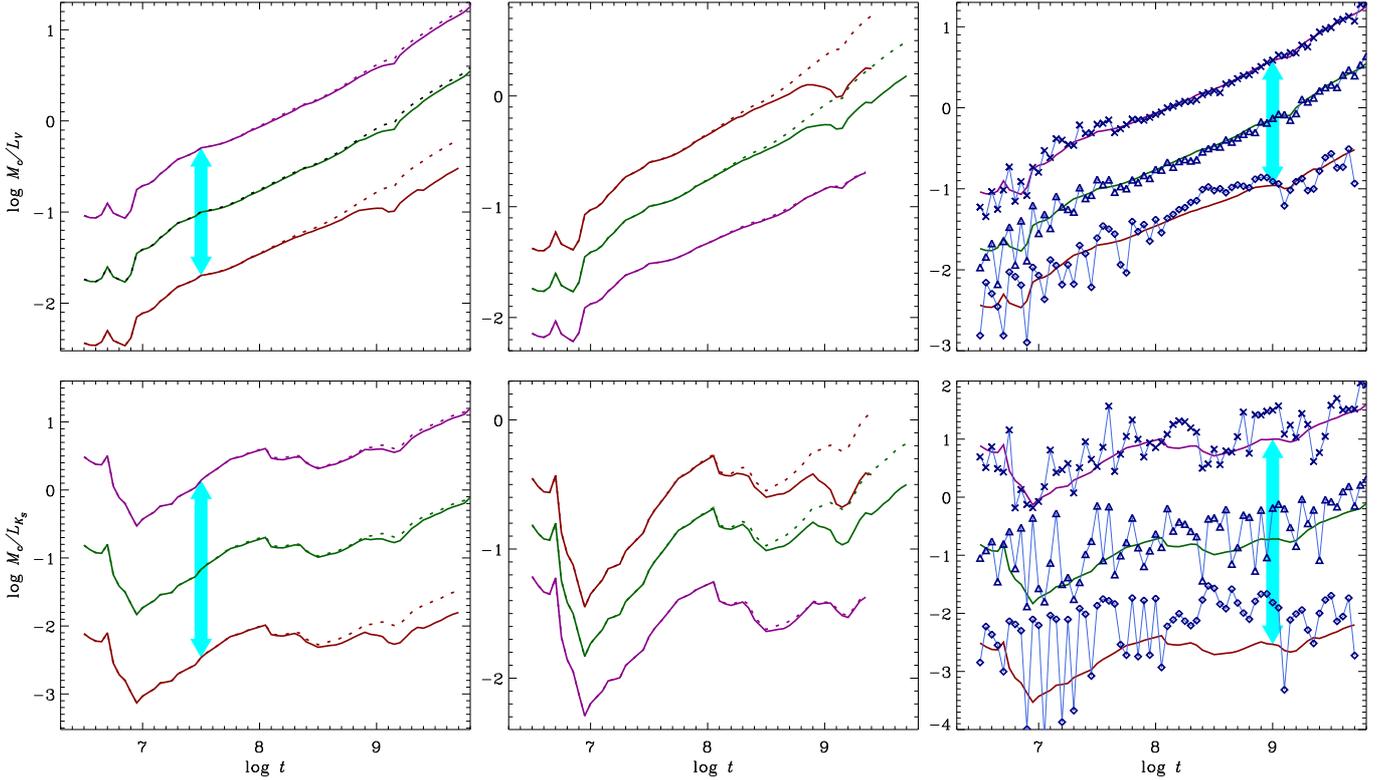}}
\caption{The evolution of the mass-to-luminosity ratio in the optical (upper row) and NIR (lower row) and various population effects. The curves shown in the left and middle panels present SSP models with continuous IMF, in the right panels we compare continuous-IMF and discrete-IMF cases. \textbf{The left panels} show the influence of cluster mass loss due to dynamical evolution for clusters of different initial masses with stellar low-mass limit $m_l=0.1\,m_\odot$. The curves are artificially shifted vertically in order to avoid crowding as it is shown by cyan arrows. The upper (shifted up) curve corresponds to the initial mass of $M_c=10^4\,m_\odot$, the middle (shown in the original position) curve is for $M_c=5\cdot 10^3\,m_\odot$, and the lower (shifted down) curve corresponds to $M_c=10^3\,m_\odot$. The dotted portions (as everywhere in this figure) mark models without mass loss. \textbf{The middle panels} show the effect of variations of the lower mass limit of stars for a model of $M_c=10^3\,m_\odot$. The upper, middle, and bottom curves correspond to $m_l=0.01,\,0.1,\,1\,m_\odot$, respectively. \textbf{In the right panels} we compare models with continuous IMF (curves) and discrete IMF (connected symbols) for clusters of different masses, having $m_l=0.1$. Crosses indicate model with $M_c=10^4\,m_\odot$, triangles $M_c=5\cdot 10^3\,m_\odot$, and diamonds $M_c=10^3\,m_\odot$. As in the left panel, the curves for different $M_c$ are artificially shifted vertically in order to avoid crowding.
}
\label{mlef_fig}
\end{figure*}
%----------------------------------------------------------------------------%

\subsection{Empirical relation}\label{sec:mlob}

\begin{figure*}[t]
\sidecaption
\resizebox{12cm}{!}{\includegraphics[clip]{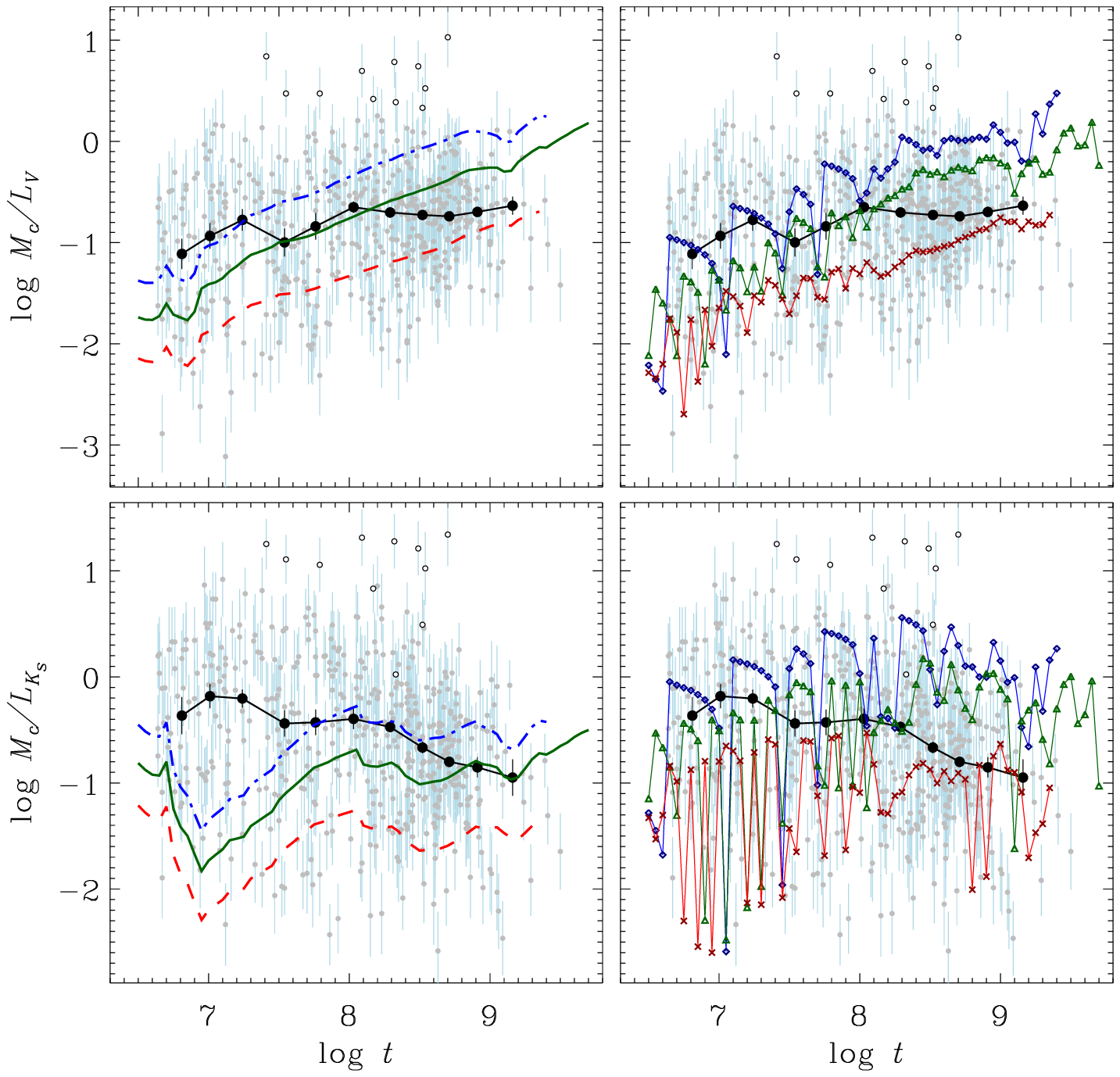}}
\caption{Comparison of observed and theoretical $M_c/L,\log t$-relations for the cases of continuous-IMF (left panels) and discrete-IMF models (right panels), cluster mass of $M_c=10^3\,m_\odot$, and variable lower limit of stellar mass $m_l$. The upper row corresponds to the optical, the lower one to the NIR passbands. The designations for the empirical data are the same as in Fig.~\ref{mlobs_fig}. For the models dash-dotted blue lines (diamonds in the right panels) correspond to $m_l=0.01\,m_\odot$, solid green lines (triangles) to $m_l=0.1\,m_\odot$, and dashed red line (crosses) to $m_l=1\,m_\odot$.
}
\label{mlcom_fig}
\end{figure*}

The sample used in this section consists of clusters with sufficiently accurate masses (error $\delta_{M_c}$ less than 80\%) from \citet{clumart,clumart1}. Further, we exclude nine objects, frequently classified as compact associations (Cep~OB3, Cyg~OB2, NGC~869,NGC~884, Nor~OB5, Sco~OB4, Sco~OB5, Sgr~OB7,  Vel~OB2) from consideration. The distribution of the remaining clusters in the $M/L$-age diagram is shown in Fig.~\ref{mlobs_fig}. A few of these clusters seem to have exaggerated values of the $M/L$ ratio and form in Fig.~\ref{mlobs_fig} a separate group of 11 objects with $M/L_V>2$, typical of globular clusters. This group consists of nearby clusters most of which are poor and irregular (Cr~65, Platais~6, Platais~9, and Turner~5), but also includes the $\alpha$~Per cluster, which can be regarded as an association, and two rich clusters Blanco~1, and Stock~2. We can not exclude that we overestimate their tidal radii, and hence their tidal masses. Therefore we mark this group with open circles in Figs.~\ref{mlobs_fig} and \ref{mlcom_fig} and do not consider them in the following discussion. The remaining sample includes 493 clusters altogether.

%----------------------------------------------------------------------------%
%----------------------------------------------------------------------------%

The straight line in the $M/L_V$-age diagram shows the empirical relation constructed from a sample of 138 clusters by \cite{battin94}. They used stellar counts for evaluating the cluster mass. Therefore, the derived masses do not take into account unseen stars and can be regarded as lower cluster mass estimates only. On the other hand, the cluster luminosity is defined by the brightest stars. It is clear that such an approach provides only a lower limit for the $M/L$-ratio. This is well-illustrated by Fig.~\ref{mlobs_fig}.

The thick curves connecting the large black dots in Fig.~\ref{mlobs_fig} represent the running median of the $M/L$ ratio with a ($\log t$)-bin of 0.5 and a step of 0.25, and it is constructed on the basis of the sample of 493 Galactic open clusters described above. Figure~\ref{mlobs_fig} shows that in the considered range of ages the $M/L_V$-age relation is a rather shapeless cloud of data points. However, a certain trend can be observed when we take the medians in age bins. We find a formal average value of $\langle M/L_V\rangle=0.26\pm0.01$ with a standard deviation of $\pm0.30$. Adopting the error of an integrated magnitude to be $\pm$0.6 (Sect.~\ref{sec:data}) and taking into account that the error in $M_c$ is better than $\pm$80\%, we derive the relative error of a single measurement of the ratio to be better than $\pm0.97$. Assuming further that the spread of points in Fig.~\ref{mlobs_fig} around the ``true'' value of 0.26 is caused solely by measurement errors, we expect a standard deviation of 0.25. This value is slightly lower than the observed scatter, and the additional dispersion produced by unknown effects should be less than 0.2.

The relation in the NIR is in many respects similar to the optical one. It also shows a considerable dispersion but there is a tendency for $M/L_{K_s}$ to decrease, as the cluster age increases. The running median of the ratio supports this impression. The formal average value of the $M/L_{K_s}$ ratio is 0.66, and the standard deviation is 1.07. The expected dispersion due to measurement errors turns out to be 0.64. This leaves much room for unknown effects (of the order of 0.86 mag).

In the sections below, we consider different SSP-models, try to fit them to the observed distribution of the $M/L$ ratio and discuss possible reasons for its relatively large scatter.

%----------------------------------------------------------------------------%
\subsection{Impact of various population effects on the $M/L$-age relation}

The effects considered in Sect.~\ref{sec:lumev}, and \ref{sec:colev}, which influence the integrated magnitudes and colours must also affect the mass-luminosity relation. As we discussed earlier the dynamical mass loss directly diminishes cluster mass. The scarcity of the population of stars in the upper Hertzsprung-Russel Diagram produces luminosity flashes of clusters. When one varies the lower mass limit of stars in a cluster with a given mass, the massive part of the IMF is changed. In this section we consider the manifestation of all these effects alone and in various combinations.

In the left panels of Fig.~\ref{mlef_fig} we show models of different masses ($M_c=10^3,\,5\cdot 10^3,\,10^4\,m_\odot$) constructed with a continuous IMF. Since the mass-to-luminosity ratio in the continuous case does not depend on cluster mass the different curves would overlap in the plot. To avoid crowding we shift (as indicated by thick arrows) the least massive model down, and the most massive model up from their original position. As one can see, the mass loss affects the later stages of the cluster evolution. At the same time clusters of different mass appear to have a different state of dynamical evolution. For example, at $t=1$~Gyr the model of $M_c=10^3\,m_\odot$ is the most dynamically evolved, and has considerably changed in the $M/L_V$ compared to the model without dynamical mass evaporation. At the same time more massive models are practically insensitive to the mass loss. Note the good agreement of our data with recent calculations of \cite{alaba09} (see their Fig.5). In spite of a considerable impact on models of smaller cluster masses, the mass loss effect is not able to change the evolutionary path of the cluster sufficiently: independent of the initial cluster mass, the $M/L$-age relation rises almost monotonically during the whole life of the cluster. Qualitatively the NIR-models demonstrate a similar behaviour as the optical ones.

In the middle panels of Fig.~\ref{mlef_fig} we show models of $M_c=10^3\,m_\odot$ with various lower limits of stellar masses $m_l$ built with a continuous IMF. They also include the effect of mass loss due to dynamical evolution. Unlike the left and right panels, the curves are at their true position without any shift. The value of $m_l$ varies from 0.01~$m_\odot$ (upper curve) through 0.1~$m_\odot$ for the middle curve, and 1~$m_\odot$ for the lower curve. Again, dotted curves correspond to no-mass-loss models, and solid curves represent evaporating models. The observed effect can be simply explained. In case of a model with $m_l=0.01\,m_\odot$, the bulk of cluster mass comes from low-mass stars, making the population of luminous stars more rare implying a lower integrated luminosity compared to the reference model of $m_l=0.1\,m_\odot$. This lead to an upward shift of the $0.01\,m_\odot$ curve with respect to the reference model. In contrast, in a model with a lower limit of $1\,m_\odot$ no low-luminous stars will be formed, and the ratio will be shifted down with respect to the reference relation. In addition, dynamical evolution will remove the bulk of cluster masses by evaporation of low mass stars. This process is more effective in clusters with low $m_l$, and it will have almost no impact on the high-$m_l$ counterpart.
 
In the right panels of Fig.~\ref{mlef_fig} we compare models of different cluster masses ($M_c=10^3,\,5\cdot 10^3,\,10^4\,m_\odot$) constructed with a continuous and discrete IMF. In order to avoid crowding, we have again shifted artificially the curves of different masses. Here we observe a picture known from the previous sections: the discreteness of  the IMF causes a spread around the continuous solutions. Whereas e.g. in $B$ we practically do not observe any influence, a bias becomes noticeable in $V$, increases considerably in the NIR, and for $K_s$ the bias between the continuous and discrete models reaches a considerable value of about one order of magnitude for clusters with $M_c=10^3\,m_\odot$. We also conclude that stochastic fluctuations in the $M/L$ ratio increase when the cluster mass decreases.

\subsection{Comparison with observations}

In Fig.~\ref{mlcom_fig} we combine the plots shown in Fig~\ref{mlobs_fig} and Fig~\ref{mlef_fig}. We plot only the subset of models that gave the best fit to the data in the optical and NIR. The curves show the models for the cluster mass $M_c=10^3\,m_\odot$ with different low-mass limits ($m_l=0.01,\,0.1$ and $1\,m_\odot$). The effect of mass loss due to the dynamical evolution is included in the models. By varying $m_l$, the continuous models can partly cover the region occupied by observations in the optical. An additional scatter is introduced,  especially for younger clusters, if one considers models with a discrete IMF. However, the main reason for a large spread of the observed $M/L_V$ relations in Fig.~\ref{mlcom_fig} is the uncertainty of individual data points (see Sect.~\ref{sec:mlob}). The typical evolution of the $M/L$ ratio in the optical can, therefore, be better understood  if we consider the evolution of the median values in  Fig.~\ref{mlcom_fig}. For the youngest clusters, the best fit is achieved by evolutionary tracks with extremely low $m_l$-limit. The clusters with $7.25 < \log t < 8.25$ are well described by the model with $m_l=0.1\,m_\odot$,  whereas the data for older clusters agree with the track with $m_l$ between $0.1\,m_\odot$ and $1\,m_\odot$.  We interpret this as a consequence of the mass loss in open clusters: the young clusters are formed with stars filling the full range of stellar masses, then, on a time scale comparable to the relaxation time, they start to evaporate (losing preferentially low-mass stars), and at older ages their effective low-mass limit sufficiently increases. This process appears to be even more dramatic than that predicted by the used N-body simulations we used to describe the dynamical evolution of open clusters. Probably, events like encounters with molecular clouds impact the stellar content of older open clusters in a very effective way.

It is evident from Fig.~\ref{mlcom_fig} (upper-left panel) that the continuous models are not able to fit the observations in the NIR: the models predict ratios too low compared to the observations, especially for younger clusters. The situation is improved if one considers models with a discrete IMF in the upper-right panel of Fig.~\ref{mlcom_fig}. In this case, one can recognise a tendency similar to that we observe for the optical passband. Young clusters tend to be fitted better by models with low $m_l$-limit, whereas evolved clusters are described more adequately by the tracks with high $m_l$-values. In contrast to the optical band, this picture is disturbed by strong fluctuations of the $M/L$ ratio towards low values due to the discreteness of  the IMF. Their contribution to the observed spread of data points is comparable with the scatter introduced by randomly distributed errors of observation.

We conclude that the evolution of the observed mass-to-luminosity ratio indicates a mass loss in open clusters with increasing age. In the optical, the large dispersion of $M/L$ ratio is caused mainly by uncertainties of the data, though the IMF-discreteness favours the observed scatter (the effect increases with decreasing cluster mass). In the NIR, the IMF-discreteness is an important contributor to the $M/L$ dispersion. The higher sensitivity of the NIR fluxes to the number of red giants in clusters explains why the NIR-diagrams show a decreasing trend with cluster age.

\section{The flashing population of local clusters}\label{sec:flashpop}

As we show throughout this paper, the discreteness of the stellar IMF plays an important role in the description of the integrated photometric properties of open star clusters and defines the flashing activity of the cluster population. In this section we try to formalise the ``discreteness term'' by deriving a condition for an emerging red flash in a star cluster. On the basis of this condition we also illustrate how the flashing activity of the cluster population depends on cluster parameters such as $M_c$ and $m_l$.

From  Fig.~\ref{colobs_fig} (left panel) we conclude that ``RG events'' are prominent in clusters younger than $\log t\approx 8.6$ where the MS-branch of the colour-age relation can be clearly recognised. Beyond that age, the  MS-branch changes its slope and transforms to the RG-branch i.e. contains mainly clusters with red giants among the brightest stars. Therefore, in the following discussion  we consider only relatively massive stars, which have a chance to become a red (super)giant during the first 400~Myr ($\log t < 8.6$), i.e. have masses above $\sim 3\,m_\odot$.

Below we construct the condition for the appearance of a red flash event: when a cluster becomes considerably brighter and redder than in the Main Sequence stage, and after a certain rather short red phase returns  back to the previous stage. This event is related to the transformation of the most massive star of a  cluster to a red giant, and can be described in terms of stellar evolution.

For the red giant flash to be noticeable, the total (MS $+$ post-MS) lifetime of the most massive star in a cluster must be shorter than the Main Sequence lifetime of the next massive star. Otherwise, the second star simply joins the first one in the Red Giant domain, and the cluster remains red. This condition allows a cluster after a post-MS flash to become a MS-cluster again, as long as the second star still stays on the MS. In other words, if the most massive star has a mass $m+dm$, and the second massive star has a mass $m$ this condition can be written as
\[
\tau_{MS}(m) > \tau(m+dm)\,,
\]
where $\tau$ and $\tau_{MS}$ refers to the total lifetime and the main sequence lifetime of a star of given mass, respectively, and $dm\ll m$. The total and Main Sequence lifetimes differ by the lifetime in the Red Giant stage $\tau=\tau_{MS}+\tau_{RG}$ (we assume that the RG-stage accumulates all the later phases of stellar life), or $\tau_{MS}=(1-\lambda)\,\tau$, where $\lambda=\tau_{RG}/\tau$. As evolutionary calculations show (see e.g. \citealt{bertelli94,girarea00,margi07,marigo08}), $\lambda$ is a weak function of stellar mass with an approximate value of $\lambda\approx 0.3\dots0.1$ in the mass range $\log m/m_\odot=0\dots1.9$. Expanding the right-hand side of the inequality above into a Taylor series, one gets
\[
(1-\lambda)\,\tau(m) > \tau(m)+\tau^\prime(m)\,dm+\dots
\]
Following stellar evolutionary calculations (e.g. Padova models),  the lifetime can be expressed as
\[
\tau(m) = \beta\,m^{-\gamma}\,,
\]
where $\gamma>0$ varies in the mass range $\log (m/m_\odot)=0\dots1.9$ from 3.8 to 0.3. Since $dm$ is relatively small, we can assume, for simplicity, the parameters $\beta,\,\gamma$ and $\lambda$ to be constant. Then the derivative can be computed as
\[
\tau^\prime(m) = -\tau(m)\frac{\gamma}{m}\,,
\]
and considering the linear terms only, we express the inequality as
\begin{equation}
 \frac{dm}{m} > \frac{\lambda}{\gamma}\,, \label{eq:flashcond}
\end{equation} 
or in other terms
\[
d\log m > \frac{\lambda}{\gamma}\,\log e\,.
\]
This supports our previous conclusions in Sect.5 that the more massive the cluster is, i.e. the denser the cluster isochrone is populated with stars ($d\log m$ being small) the harder it is for the cluster to produce the flashing. This means, also, that the ability of a cluster to demonstrate red flashing is controlled by stellar evolution parameters: the duration of the red-giant phase $\lambda$, and the decrement of the lifetime $\gamma$. The more pairs of stars having masses $\log m/m_\odot >0.5$ will fulfil the condition (\ref{eq:flashcond}) at birth, the more RG-flash events a cluster will undergo in its life. We use this property to build a simple model of cluster population below.

%----------------------------------------------------------------------------%
\begin{figure}[t]
%\sidecaption
\resizebox{\hsize}{!}{\includegraphics[clip]{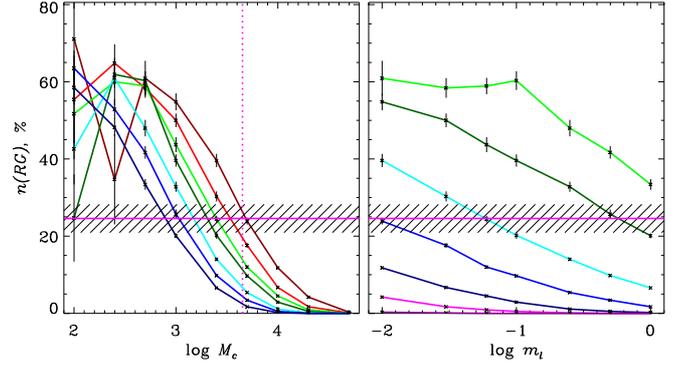}}
\caption{The dependence of frequency $n(RG))$ of red flashes on cluster parameters as derived from Monte-Carlo experiments. The left panel shows a dependence on $M_c$ with $m_l$$=$0.01, 0.03, 0.06, 0.1, 0.25, 0.5,1 $m_\odot$ (from top to bottom along the dotted line). The right panel shows a dependence of $n(RG)$ on $m_l$ with $M_c$$=$$5\cdot 10^2$, $10^3$, $2.5\cdot 10^3$,$5\cdot 10^3$, $10^4$, $2\cdot 10^4$, $5\cdot 10^4$ $m_\odot$ (from top to bottom). The bars indicate  statistical uncertainties derived in the experiments. The horizontal solid line marks the observed frequency of local RG-clusters within the completeness area of the cluster sample of 850 pc. The hatched area corresponds to the Poisson error of this estimate. The vertical dashed line in the left panel shows the estimated value of the average mass of formed clusters from \citet{fuma}.
}
\label{nrg_fig}
\end{figure}
%----------------------------------------------------------------------------%

We stress that the condition (\ref{eq:flashcond}) does predict only the case of the colour bias with respect to the continuous SSP and does not deal with colour fluctuations around the SSP that can occur even if (\ref{eq:flashcond}) does not hold. Thus, we consider this condition as a colour-bias constraint and apply it below to reproduce the flashing statistics of the local cluster population. We apply the colour-bias condition to a mass spectrum of a newly formed cluster constructed by a numerical simulation as described in Sect.~\ref{sec:rand}. The basic parameters of the simulation are cluster mass $M_c$, the lower limit of stellar masses $m_l$, and the slope $\alpha$ of the IMF (assumed to be 2.35 across our study). These model parameters define the population of the cluster and in particular, at masses greater than a limit of  $\log m_0/m_\odot =0.5$ adopted above. For each pair of stars in this mass range, we compute the difference in $\log m$ and apply the condition (\ref{eq:flashcond}) to them. Those stars which satisfy this condition are marked as flash-producing. In order to get statistically reasonable results, we repeat this procedure several times (usually a few tens of simulations are sufficient to get statistically meaningful results). As we noted above, we do not need to follow the whole evolution of a cluster to get the flashing-activity history. It is sufficient to analyse the initial mass spectrum only.

The results of the simulations are presented in Fig.~\ref{nrg_fig} which shows the RG-flash frequency (the ratio of the number of stars satisfying condition (\ref{eq:flashcond}) to the total number of stars formed with $m>m_0$) for different $M_c$ and $m_l$.  At $M_c>500\,m_\odot$ the curves show a regular pattern: the RG-flash frequency decreases with increasing  $M_c$. At $M_c=5\cdot 10^4\,m_\odot$, the cluster isochrone is so densely populated with massive stars that the RG-flashes are practically no longer possible at either value of $m_l$. In contrast, at cluster masses smaller than $M_c=500\,m_\odot$, the flashing activity of clusters is constantly high ($n(RG)\sim50\dots60\%$), but the isochrone is so poorly populated that the behaviour of the model curves is rather volatile. In the regular part of the diagram, one sees a diminution of the frequency $n(RG)$ at given $M_c$ with increasing $m_l$. Again, this occurs due to a rise of the stellar density of stars with $\log m_0/m_\odot > 0.5$ in the corresponding range of the isochrone.

We also show in Fig.~\ref{nrg_fig} the empirical estimate of the flashing frequency in the Solar Neighbourhood which is found to be $25\pm4\%$. The estimate is derived from our \ascc sample of open clusters which are younger than $\log t=8.6$. To avoid a selection effect, we involve only clusters residing within 850 pc from the Sun. According to \citet{clupop}, this distance corresponds to the completeness distance of our sample, and guarantees that at smaller distances we are free from the incompleteness bias. The estimate was computed as the ratio of the number of clusters having red giants (45) to the total number of selected clusters (183). The empirical value of $n(RG)$ can be somewhat overestimated, since we are not sure that all observed RG-clusters undergo currently a flash (i.e. they may come back to the MS state after the observed RGs will die). However, as a careful inspection shows, most of the nearby clusters are ``flashing-biased'', i.e. are poor enough and contain only a few red giants. In the right panel of Fig.~\ref{nrg_fig} we show the $n(RG)-m_l$-relation with $M_c$ taken as a parameter. One can see that the observations are compatible with initial cluster masses in the range $M_c=10^3\dots5\cdot 10^3\,m_\odot$.

In principle, Fig.~\ref{nrg_fig} shows how statistics on the red giants (the brightest stars of a cluster) can be converted into a knowledge on the lower mass limit of forming stars provided that the initial cluster mass is known. The typical mass of young clusters of $4.5\cdot 10^3\,m_\odot$ we found from the analysis of the cluster mass function  \citep{fuma} indicates that $m_l$ should be small. Taking into account the  uncertainty both in the frequency, and in the initial mass values, we expect a possible $m_l$ range of $0.01\dots0.1\,m_\odot$.

\section{Summary and Conclusions}
\label{sec:conc}

We have proved the necessity to modify the traditional SSP-model widely applied in studies of the extragalactic clusters because it fails to explain the luminosity and colour distribution and evolution of Galactic open clusters \citepalias{clussp}. The observed colour offset between theory and observation led us to the idea to revise the basic assumption used in  present-day SSP -- the continuity of the IMF of cluster stars. The successful solution of the observed discrepancy in \citetalias{clussp} provided an impetus for the extension of the investigation of the properties of the discrete SSP model. This paper includes the analysis of the evolution of luminosity, colour, and the mass-to-luminosity ratio with cluster age of both continuous and discrete SSP-models. The study includes several effects which either occur only in a discrete approximation (like the $m_l$-effect in cluster colours), or were rarely studied even in the continuous approximation (mass loss from clusters).

The evolution of the integrated magnitudes shows that  mass loss is the weakest effect among all others taken into account. It caused only a small change of cluster magnitude of older clusters. This can be expected since in the considered regime, mass loss occurs mainly due to evaporation of the low-mass stars that only weakly contribute to the cluster luminosity. However, the mass loss effect taken into account modifies the general behaviour of the evolutionary track, increases the steepness of the magnitude-age relation, and thus provides a better agreement with the observations. The effect can possibly become more prominent when the other causes of mass loss (which are not considered here) are taken into account, e.g. encounters of clusters with molecular clouds.

The effect of the lower limit of stellar masses $m_l$ is stronger. Being applied to a cluster of a particular mass it alone is able to account for the total width of the observed magnitude-age relation. Unlike the mass loss it does not affect the slope of the relation.

Last but not least, the IMF-discreteness effect has large consequences for the cluster luminosity. It is responsible for cluster flashing and track fluctuations due to the appearance of red giants. Moreover, it is the reason for a systematic bias between luminosities of MS-clusters and the predictions by standard SSP-models. The effect is especially strong at long wavelengths, where the flash amplitude can reach  4 mag, and the branch of MS-clusters may be located 2-3 magnitudes below the continuous track.

In summary, we conclude that the general evolution of Galactic open clusters in the magnitude-age diagram is adequately described by models of moderate mass ($5\cdot 10^2\dots$ a few of $10^3\,m_\odot$). We explain the difference between this value and the average mass of forming clusters of $4.5\cdot10^3\,m_\odot$ \citep{fuma} by the episodes of cluster mass loss \citep{weidea07}, which occurs in young clusters as a result of violent events (SN explosions, HII formation, and stellar winds), by mass loss due to encounters with molecular clouds, and due to regular evaporation.

All this indicates that the approximation of the observed relations with a simple grid of continuous SSP-tracks is much too simplified. For example, the extragalactic cluster samples can be well contaminated at magnitudes $M_V=-6\dots-10$ mag with flashing clusters, which in reality have lower MS-luminosity and cluster mass. This in turn can make the observed sample artificially overpopulated by luminous and massive clusters and will make the apparent luminosity and mass functions flatter.

Therefore, the direct implementation of standard SSP models, as it is usually applied to extragalactic clusters to estimate their masses, must be taken with caution. The only possibility to avoid large errors in mass estimates is using observations at short wavelengths (e.g., $U$, $B$).

Cluster colours also stress (sometimes even more than the magnitudes do) the  trends mentioned above. For example, the discrete models demonstrate a colour-dependence on the low-mass limit, or on $M_c$. Sometimes we see that both effects mimic each other in the discrete case. This is impossible in the case of models with a continuous IMF where the cluster colours are, by definition, independent of $M_c$ and $m_l$.

Again the basic difference between the standard and discrete models is the colour bias produced by MS-clusters, and the flashing activity related to the formation of red (super)giants. As for integrated magnitudes, the prominence of both events depends on the population of the massive end of the stellar mass function in a cluster and is directly related to $M_c$ and $m_l$. As shown in \citetalias{clussp}, with an average value in $\Delta (B-V)\approx0.3$ mag the colour bias reaches its maximum of the order of 0.5 mag at $\log t\approx 7$ and can be still detected at $\log t>8$. Again, the colour bias increases at longer wavelengths and reaches a value of the order of 1 mag in $(J-K_s)$. The presence of the bias in the standard SSP-models should be taken into account by determining cluster reddening and age from the integrated colours of clusters with masses less than $10^6\,m_\odot$, and warns against a straightforward use of the SSP-grids \citepalias{clussp}.  The discrete model with a cluster mass of $10^3\dots$ is successful in reproducing the colours of young clusters ($\log t < 8.5$) of our cluster sample. At older ages the flashing population is represented mainly by clusters of lower masses with different evolutionary stage of red giants along the RG-sequence.

It is evident that an independent approach for the determination of reddening and mass is required before cluster colours can be used for the age estimate. But even in this case, the discreteness effect must be taken into account, and the age will be determined within some confidence interval by interpolating a grid of a properly selected  SSP-model. In order to increase the reliability of the results, the age determinations should be based on observations in short wavelengths.

We have constructed the mass-to-luminosity ratio based on luminosity and  mass data determined independent of each other for Galactic open clusters. This makes the derived relation $M/L$-age  bias free. The average relation increases with age in the optical, and decreases in the infrared. The observed scatter turns out to be slightly larger than that expected from the accuracy of the input values in the optical, but considerably larger in the infrared.

If we introduce the effects described above in our models, the following picture  emerges. Mass-loss in clusters makes the $M/L$ tracks to appear flatter at later stages ($\log t>8.5$). The less massive the cluster is, the more pronounced is the effect. The combination of mass loss and  $m_l$ makes the curves almost flat in the NIR after $\log t\approx 8$. The variations of $m_l$ within the  limits adopted in this study are able to produce a considerable vertical scattering in  the relation, and the IMF-discreteness adds extra fluctuations. In the NIR at smaller cluster masses one observes a strong bias in the relation, implying that the continuous option predicts too low $M/L$.

Compared to the observations, the tracks show better agreement with a discrete IMF, covering practically the complete observed $M/L$ spread. One also notes an interesting detail, the younger $(\log t<7.5)$ clusters are in better agreement with tracks with lower values of $m_l$, and older clusters ($\log t>8.5$) agree with higher-$m_l$ tracks. We interpret this finding as an indirect evidence of the dynamical evolution of open clusters, born with a full mass spectrum of stars, which is narrowed from the massive end by stellar evolution, and from the low-mass end by dynamical evolution. Probably, the mass loss due to irregular events (e.g., encounter with molecular clouds) may play an important role on the decreasing of the $M/L$-ratio in older clusters.

Taking into account the time-scales of stellar evolution, we derived a condition for the occurrence of colour and magnitude flashes observed in the photometric diagrams for open clusters. This condition reflects well the changing properties of stars caused by their evolution. Unlike the approach by \citet{cervino04} to separate physically feasible and un-realistic SSP-models, our condition reveals the origin of a bias in realistic models. Our model appears to be very convenient in predicting the flashing activity of clusters. We use this ``flashing condition'' for the construction of a simple population model of star clusters, which relates the frequency of RG-flashes, cluster mass, and low-mass limit of cluster stars. The application of this model to open clusters in the Solar Neighbourhood has the following consequences: with the average observed frequency of flashing clusters of 25$\pm$4\,\% and with an average mass at birth of a few $10^3\,m_\odot$, we infer that in newly formed clusters the lower limit of the stellar mass spectrum should fall into the range $0.01\dots0.1\,m_\odot$.

\begin{acknowledgements}
This study was supported by DFG grant RO 528/10-1, and RFBR grant 10-02-91338. We thank the anonymous referee for her/his detailed comments.
\end{acknowledgements}

\bibliographystyle{aa}
\bibliography{clubib}

\end{document}